\documentstyle[psfig,12pt]{article}

\begin{document}

\sloppy

\baselineskip=20pt

\begin{center}
{\Large\bf FIRST-radio sources in clusters of galaxies}\\[8mm]
{\bf  A.G.~Gubanov, V.P.~Reshetnikov}\\[6mm]
{\it  Astronomical Institute of St.Petersburg State University, Russia}\\[15mm]
\end{center}

\begin{center}
{\bf Abstract}
\end{center}

{\small{\bf
The detailed identifications of the FIRST and NVSS radio sources
with optical objects in APM and in DSS surveys are carried out
for 26 rich Abell clusters of galaxies.
99 radio sources are identified with optical objects,
40 have probable identifications,
and 187 are not identified from 326 radio sources in the considered fields
(within Abell radius of each cluster of galaxies).
20 radio sources are definitely in clusters and 34 can be clusters members with
high confidence. Therefore, $\approx$30-40\% of the FIRST
radio sources can be identified with optical objects on the base
of APM and of DSS data. On average, one can find 2 identified
sources per cluster.}}

\vspace{1.5cm}

In the framework of previously announced project (A.G.Gubanov et al. 
1997, in ``Problems of modern radioastronomy'', XXVII Radio Astronomy
conference, St.Petersburg, V.1, P.168), we present
detailed identifications of the FIRST (and NVSS) radio sources
with objects from APM and DSS surveys
for 26 rich Abell clusters of galaxies (Table 1).

Identifications are performed for all radio sources within Abell
radius of cluster of galaxies (3 Mpc for the Hubble constant of
50 km/s/Mpc). The clusters selected within the processed zone of 
the FIRST survey are part of a larger sample of rich clusters 
(with richness $\geq$2) and of clusters containing cD-galaxies.

\newpage

\begin{verbatim}
  Table 1.   List of clusters.

 ACO  RA(J2000)  DEC(J2000)    z       ACO  RA(J2000)  DEC(J2000)    z
        h  m  s     o  '  "                   h  m  s     o  '  "
A0586 07 32 17.8 +31 37 34  0.1710    A1033 10 31 33.8 +35 04 34  0.1258
A0642 08 19 05.7 +30 02 34  0.2063e   A1035 10 32 07.3 +40 12 33  0.0785
A0690 08 39 14.3 +28 51 24  0.0788    A1068 10 40 47.2 +39 58 20  0.1375
A0705 08 47 38.7 +30 00 56  0.1138e   A1073 10 42 26.6 +36 38 17  0.1390
A0715 08 54 44.4 +35 24 33  0.1685e   A1081 10 44 49.5 +35 34 14  0.1585
A0727 08 59 13.1 +39 26 19  0.0992e   A1094 10 47 32.8 +27 31 10  0.2004
A0781 09 20 23.2 +30 26 16  0.1763e   A1120 10 53 15.5 +30 48 02  0.2218e
A0800 09 28 30.0 +37 47 53  0.2223e   A1175 11 09 13.7 +33 10 44  0.2487
A0812 09 32 35.4 +37 53 42  0.1452e   A1178 11 09 50.1 +34 35 43  0.2596
A0908 09 59 35.8 +22 25 36  0.2026e   A1182 11 10 19.0 +31 46 43  0.1660e
A0943 10 12 12.2 +33 37 10  0.1485e   A1190 11 11 46.2 +40 50 42  0.0763
A0961 10 16 29.5 +33 37 01  0.1241    A1198 11 12 48.0 +30 22 40  0.1660e
A0963 10 17 09.7 +39 01 00  0.2060    A1258 11 26 08.7 +25 26 30  0.1469e
\end{verbatim}

We obtained the following results:

1. Among 326 radio sources in the considered
fields of 26 clusters (and within Abell radius),
99 radio sources are confidently identified with optical objects in the
APM and DSS surveys, 40 sources have
probable identifications, and 187 are not identified. \\

Among identified radio sources, 20 are definitely in clusters and
34 can be clusters members with high probability. \\

Therefore, the expectations to identify $\approx$30-40\% FIRST
radio sources and to find, on average, 2 sources per cluster are
coming true. \\

One can expect to obtain $\approx$8000 identifications of the galaxies
in ACO-clusters with the FIRST-type radio sources in all Abell clusters.
\vspace{3mm}\\
2. Only 11 radio sources consist of one component (within the FIRST-survey
resolution of about 5'') and only 5 are compact (with diameter $\leq$5'')
among 20 definitely lying in clusters sources. \\

Morphology of identified radio sources is typical for clusters 
radio-galaxies: \\

  5 have WAT structure;
  
  5 -- HT;
  
  5 - extended FR~I-type radio sources (with uncertain morphological type);
  
  5 - compact.\\
  
Centers of radio sources are often essentially displaced from the positions of
host galaxies in our sample and,
therefore, visual analysis of radio-optical maps is needed for correct
identification. Graphical interface of the database of clusters created at
the Astronomical Institute of the SPbSU (A.G.Gubanov, V.B.Titov,
1997, in ``Problems of modern Radio Astronomy'', XXVII Radio Astronomy 
conference, St.Petersburg, V.1, P.320) is an excellent and useful
tool for such purposes (URL {\large{\bf
http://future.astro.spbu.ru/Clusters.html}}).
\vspace{3mm}\\
3. In most cases the identified radio sources have radio 
luminosities,
log(L$_r$ [erg/sec]), typical for radio galaxies -- 40.5-42 within
frequency diapason
10 MHz -- 10 GHz  and are identified with bright elliptical galaxies.
\vspace{3mm}\\
4. It seems interesting that dynamically active clusters (consisting
of subclusters of comparable richness) often contain several
radio galaxies with comparable power (for instance, A1033, A1035, A1068,
A1081, A1190, A1258). This is not typical for more regular clusters.
\vspace{3mm}\\
5. It should be noted that the flux densities in the
FIRST, NVSS catalogues and apparent magnitudes in the APM-survey
are contain often significant systematic and random errors and
cannot be useful for correct statistical analysis.
Careful individual reduction of the data of digitized surveys and special
measurements are highly nedeed.
\vspace{5mm}\\  
Several examples of identified radio sources are presented in Fig.1--3.
\vspace{7mm}\\
This research is supported by the Federal Programme``Integration''
(project $N$~578) and by the Russian Foundation for Basic Research
(grant $N$~97-02-18212).

\begin{figure}
\centerline{\psfig{file=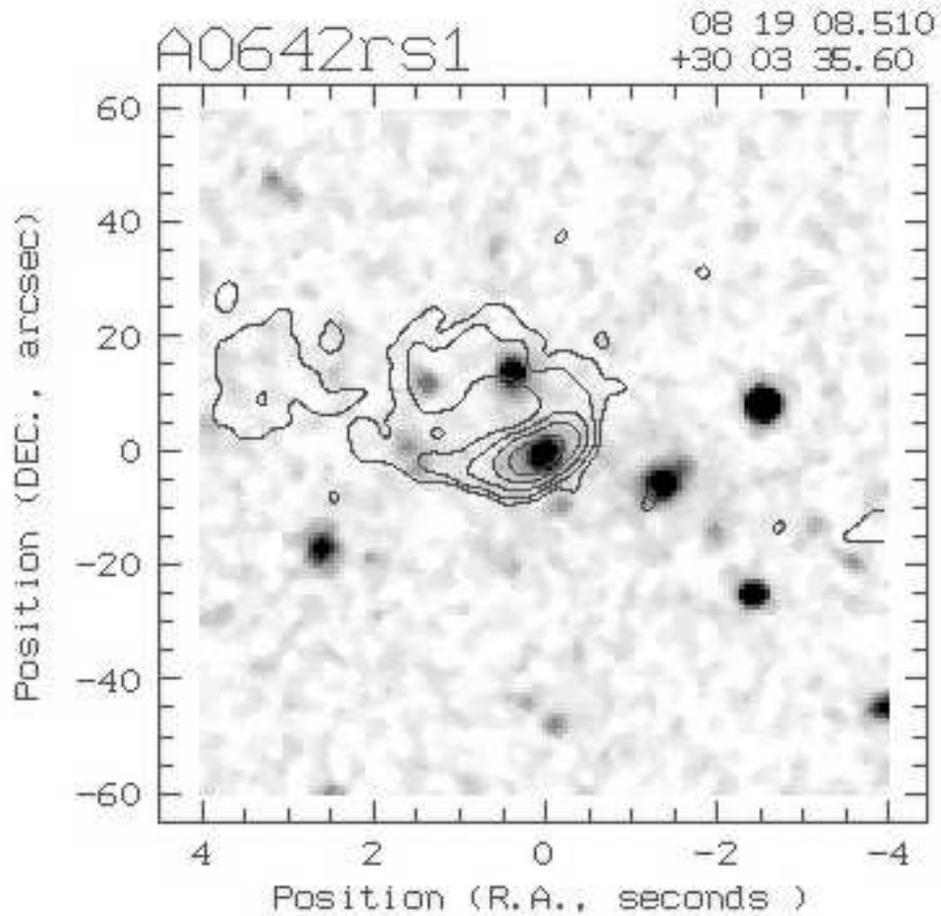,width=10cm}}
\caption[1]{One from two brightest galaxies in cluster center, 
HT radio galaxy.
Faint compact detail in the tail of this radio source
coincides in position with other cluster galaxy
(i.e. possible cluster radio galaxy too).}
\end{figure}

\begin{figure}
\centerline{\psfig{file=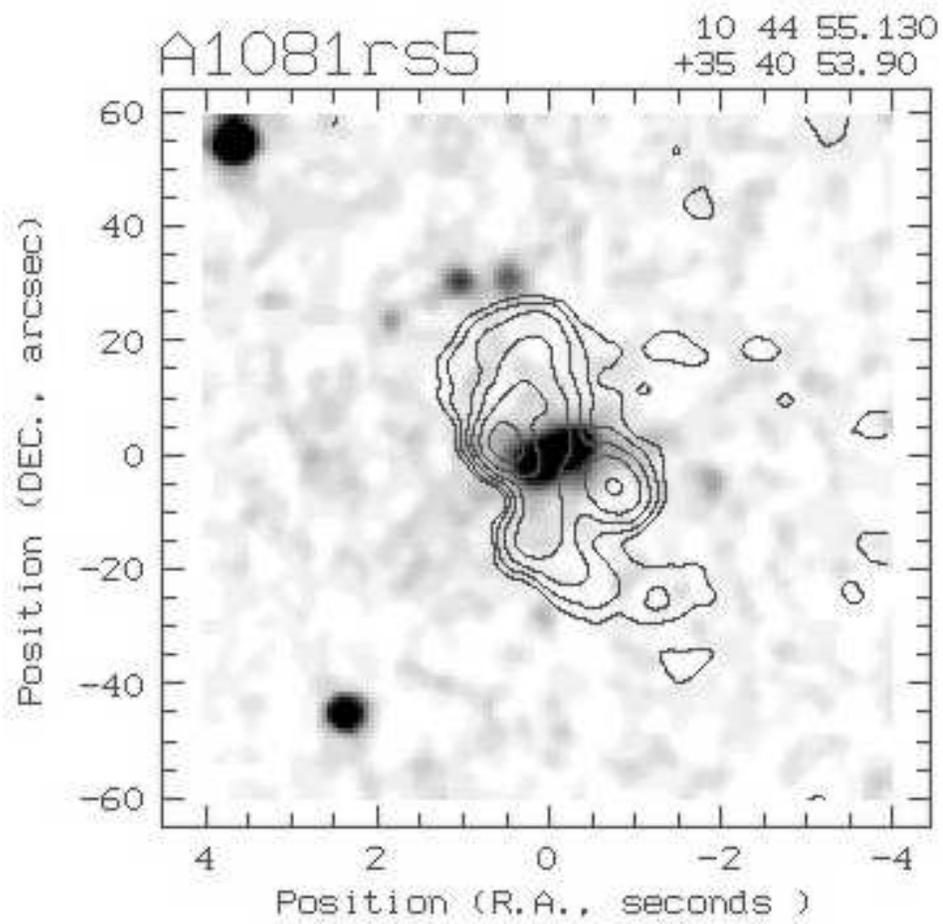,width=10cm}}
\caption[2]{The pair of close galaxies, probably, eastern is radio galaxy.}
\end{figure}

\begin{figure}
\centerline{\psfig{file=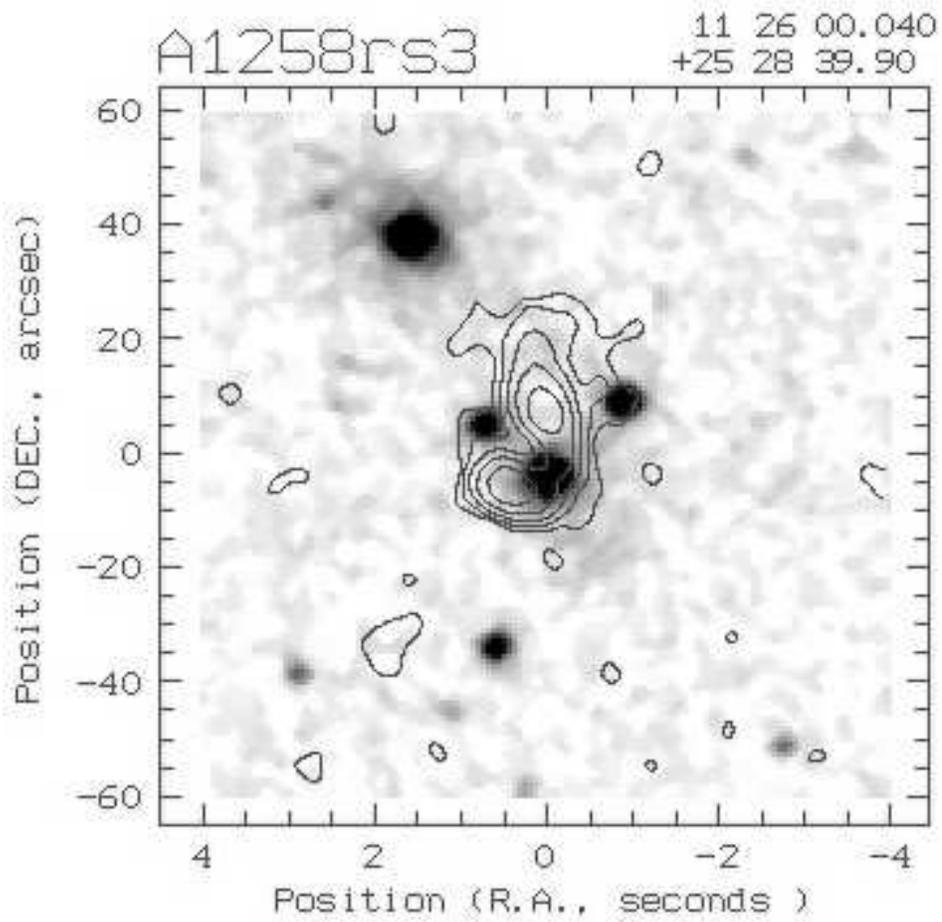,width=10cm}}
\caption[3]{HT radio source is identificated with second bright 
galaxy in cluster center.}
\end{figure}

\end{document}